\RequirePackage{fix-cm}
\documentclass[twocolumn,epjc3]{svjour3}  
\smartqed  
\RequirePackage{graphicx}
\RequirePackage{xcolor}
\RequirePackage{amsmath}
\RequirePackage{amsfonts}
\RequirePackage{float}
\RequirePackage[
            citestyle=numeric,
            pagetracker=true,
              sorting=none,
              backend=bibtex,
              doi=true,
              isbn=false,
              url=false,
              eprint=false,
              mincitenames=3,
              maxcitenames=3
              ]{biblatex} 
\bibliography{main}
\usepackage{soul}
\usepackage[normalem]{ulem}

\journalname{Eur. Phys. J. A}
\begin{document}
\sloppy
\title{Employing deep-learning techniques for the conservative-to-primitive recovery in binary neutron star simulations}

\author{Ranjith Mudimadugula\thanksref{UP}
        \and
        Federico Schianchi\thanksref{UP,UIB}
        \and 
        Anna Neuweiler\thanksref{UP}
        \and 
        Thibeau Wouters\thanksref{GRASP,Nikhef}
        \and
        Henrique Gieg\thanksref{UP}
        \and
        Tim Dietrich\thanksref{UP,AEI}
}

\institute{Institut f\"{u}r Physik und Astronomie, Universit\"{a}t Potsdam, Haus 28, Karl-Liebknecht-Str. 24/25, 14476, Potsdam, Germany \label{UP}
\and
Departament de Física \& IAC3, Universitat de les Illes Balears, Palma de Mallorca, Baleares E-07122, Spain \label{UIB}
\and
Institute for Gravitational and Subatomic Physics (GRASP), Utrecht University, Princetonplein 1, 3584 CC Utrecht, The Netherlands \label{GRASP}
\and
Nikhef, Science Park 105, 1098 XG Amsterdam, The Netherlands \label{Nikhef}
\and
Max Planck Institute for Gravitational Physics (Albert Einstein Institute), Am M\"uhlenberg 1, Potsdam 14476, Germany\label{AEI}
}

\date{Received: date / Accepted: date}

\maketitle

\begin{abstract}
The detection of GW170817, together with its electromagnetic counterparts, has proven that binary neutron star mergers are of central importance to the field of nuclear astrophysics, e.g., through a better understanding of the formation of elements and novel constraints on the supranuclear dense equation of state governing the matter inside neutron stars. 
Essential for understanding the binary coalescence are numerical-relativity simulations, which typically come with high computational costs requiring high-performance computing facilities. 
In this work, we build on recent studies to investigate whether novel techniques, such as neural networks, can be employed in the conversion of conservative variables to primitive hydrodynamical variables, such as pressure and density. 
In this regard, we perform -- to the best of our knowledge -- the first binary neutron star merger simulations in which such methods are employed. 
We show that this method results in stable simulations, reaching accuracies similar to traditional methods with an overall comparable computational cost. 
These simulations serve as a proof of principle that, in the future, deep learning techniques could be used within numerical-relativity simulations.
However, further improvements are necessary to offer a computational advantage compared to traditional methods. 

\keywords{General Relativistic Hydrodynamics\and Binary Neutron Star Mergers \and Equation of State \and Neural Networks}
\end{abstract}

\section{Introduction}
\label{sec:intro}

The multi-messenger detection of GW170817~\cite{LIGOScientific:2017vwq}, GRB170817A~\cite{Goldstein:2017mmi}, and AT2017gfo~\cite{Coulter:2017wya} has been a breakthrough in the field of multi-messenger astronomy by incorporating gravitational waves as an observational window. This groundbreaking event, associated with the merger of two neutron stars, has already provided significant insights into the evolution of the Universe~\cite{LIGOScientific:2017adf,Hotokezaka:2018dfi,Nakar:2020pyd,Coughlin:2020ozl,Dietrich:2020efo}, the properties of matter at supranuclear densities~\cite{LIGOScientific:2017vwq,Dietrich:2020efo,Annala:2017llu,Bauswein:2017vtn,Fattoyev:2017jql,Ruiz:2017due,Radice:2017lry,De:2018uhw,Most:2018hfd,Capano:2019eae,LIGOScientific:2018hze,LIGOScientific:2018cki,Coughlin:2018miv,Coughlin:2018fis,Huth:2021bsp}, and the processes responsible for the formation of heavy elements~\cite{Cowperthwaite:2017dyu,Smartt:2017fuw,Kasliwal:2017ngb,Kasen:2017sxr,Watson:2019xjv,Rosswog:2017sdn,LIGOScientific:2017pwl}.

To extract information from such high-energetic astrophysical observations, one has to compare the measured data with predictions of the binary evolution. Regarding gravitational-wave (GW) astronomy, this can be done by correlating the measured strain signal with existing GW models, maximizing their agreement~\cite{Veitch:2014wba}. For an interpretation of possible electromagnetic (EM) counterparts, one needs to relate the properties of the observed signals with the predicted light curves and spectra caused by the material outflow from two neutron stars during the coalescence~\cite{Radice:2018pdn,Waxman:2017sqv}. Because of the complexity of the merger process, an accurate description of the system's dynamics requires us to solve Einstein's field equations together with the equations of general relativistic hydrodynamics. For this reason, we have to perform numerical-relativity (NR) simulations to understand the binary neutron star (BNS) merger dynamics. In general, such simulations also allow us to extract the emitted GW signal and to validate/calibrate existing GW approximants, e.g.,~\cite{Dietrich:2017aum,Dietrich:2017feu,Kawaguchi:2018gvj,Abac:2023ujg,Gamba:2023mww}. In addition, they also allow us to understand the properties of the outflowing material~\cite{Radice:2018pdn, Hotokezaka:2012ze, Nedora:2020qtd}. Regarding the latter, one requires a proper description of microphysical processes.
This implies that one has to use a nuclear-physics motivated equation of state (EOS) for the description of the supranuclear dense material, incorporate a description of neutrino radiation and of magnetic fields; cf.~e.g., ~\cite{Rosswog:2003rv,Liebling:2010bn,Sekiguchi:2011zd,Abdikamalov:2012zi,Mosta:2013gwu,Galeazzi:2013mia,Giacomazzo:2014qba,Kiuchi:2015sga,Foucart:2017mbt,Kiuchi:2017zzg,Foucart:2018gis,Palenzuela:2018sly,Foucart:2020qjb,Radice:2021jtw,Schianchi:2023uky} for such efforts. 

Unfortunately, even by employing approximations for the microphysical description, NR simulations come with high computational costs, and single simulations require between tens of thousands to several millions of CPU hours, depending on the incorporated physical descriptions and the employed resolution. Hence, we can typically only investigate the last few orbits and the first tens to hundreds of milliseconds after the merger; cf.~\cite{Kiuchi:2022nin, Ng:2023yyg} for recent state-of-the-art simulations covering up to one second.

This large computational footprint motivates the continuous effort within the NR community to increase the efficiency of NR simulations. While there has been progress, e.g., through the application of new discretization schemes, e.g.,~\cite{Bugner:2015gqa,Kidder:2016hev,deppe_2022_6127519,Adhikari:2025nio}, or the usage of GPUs, e.g.,~\cite{Shankar:2022ful,Zhu:2024utz}, some works also investigated possibilities to speed up individual routines such as the conservative-to-primitive recovery~\cite{Galeazzi:2013mia, Kastaun:2020uxr,Dieselhorst:2021zet}. The conversion from the evolved conservative variables to the primitive variables is a critical ingredient within the often employed Valencia formulation of general-relativistic hydrodynamics~\cite{Font:2008fka, Anton:2005gi}.  
In this article, we want to build on the previous works of Refs.~\cite{Dieselhorst:2021zet, Kacmaz:2024fwa,Wouters} and present, to our knowledge, the first general-relativistic simulations in which deep-learning techniques are employed to speed up the conservative-to-primitive recovery during BNS simulations. 

The article is structured as follows. In Sec.~\ref{sec:method} we will review the basic methods and also discuss the constructed neural networks (NNs). In Sec.~\ref{sec:validation} we perform validation tests of our new method and in Sec.~\ref{sec:BNS} we present results of BNS simulations. Throughout this article, we will use geometric units $G = c = M_\odot = 1$ for all our calculations, which means that one code unit for lengths refers to $\simeq1.48\rm km$, for times refer to $\simeq5\mu s$, and for masses refer to $1M_\odot$.  

\section{Numerical Methods}
\label{sec:method}

\subsection{General relativistic hydrodynamics simulations}
\label{sec:GRHD_equations}
The evolution equations to describe general-relativistic hydrodynamics for a perfect fluid, neglecting the influence of magnetic fields or neutrino radiation, can be derived from the conservation laws of the baryon number and the energy-momentum:
\begin{equation}
    \nabla_\mu \left(\rho u^\mu\right) =0, \hspace{0.5cm} \nabla_\mu T^{\mu\nu} = 0.
\end{equation}
   
To write the evolution system into the form of a balance law
\begin{equation}
    \frac{\partial \textbf{q}}{\partial t} + \frac{\partial \textbf{F}^i}{\partial x^i} = \textbf{s},
    \label{eq:balancelaw}
\end{equation}
one typically transforms the primitive variables $\textbf{w}=\left(\rho, v^i, \epsilon, p \right)$, corresponding to the rest-mass energy density $\rho$, the fluid velocity $v^i$, the specific internal energy $\epsilon$, and the pressure $p$, to the following conservative variables $\textbf{q}$:
\begin{align}
    D    =& \sqrt{\gamma} \rho W,\label{cons_D} \\
    S_j  =& \sqrt{\gamma}  \rho h W^2 v_j,\label{cons_S}\\
    \tau =& \sqrt{\gamma}  \left(\rho h  W^2 - p \right)- D,\label{cons_tau} 
\end{align}     
corresponding to the rest-mass density $D$, the momentum density $S_j$, and internal energy $\tau$ as seen by Eulerian observers.
Here, $\gamma$ is the determinant of the induced three-metric $\gamma_{ij}$, $W$ is the Lorentz factor, and $h=1+\epsilon + p/\rho$ is the fluid's specific enthalpy. Note that the total mass-energy density as measured
by an observer comoving with the fluid is $\rho^* = \rho(1 + \epsilon)$.

This choice leads to the following evolution equations in the form of Eq.~\eqref{eq:balancelaw} with
\begin{align}
   \textbf{q}   &= [D, S_j, \tau], \\ 
   \textbf{F}^i &= \alpha \begin{bmatrix} 
                       D \tilde{v}^i \\ 
                       S_j \tilde{v}^i + \sqrt{\gamma} p \delta_j^i \\ 
                       \tau \tilde{v}^i + \sqrt{\gamma} p v^i  \\
                    \end{bmatrix}
   \label{fluxterm}, \\               
   \textbf{s} &= \alpha \sqrt{\gamma} \begin{bmatrix}  
                    0 \\
                    T^{\mu \nu} \displaystyle \left( \frac{\partial g_{\nu j}}{\partial x^\mu} - \Gamma^\lambda_{~\mu \nu} g_{\lambda j} \right)\\ 
                    \alpha \left(T^{\mu 0} \displaystyle \frac{\partial \ln (\alpha)}{\partial x^\mu}-T^{\mu \nu} \Gamma_{~\mu \nu}^0\right)\\                       
                    \end{bmatrix}.
\end{align} 
Here, we introduced $\tilde{v}^i = v^i -\beta^i/\alpha$, with $\alpha$ the lapse function and $\beta^i$ the shift vector~\cite{Alcubierre:2008}, $T^{\mu \nu}$ the stress-energy tensor, and the Christoffel symbols $\Gamma_{~\mu \nu}^\lambda$.

To finally close the system of equations, considering the existence of five conservative variables for which evolution equations exist and the presence of six primitive variables, one requires an EOS that relates, for example, the pressure to other hydrodynamical variables. In this article, we are employing two different EOSs: (i) an ideal gas EOS in which the pressure is given by $p=(\Gamma-1)\rho \epsilon$ and (ii) a tabulated EOS representing a microphysical EOS for which the pressure depends on the temperature $T$, the baryon number density $n_B$, and the electron fraction $Y_e$. For simplicity, we restrict our studies to the relativistic mean-field EOS SFHo~\cite{Steiner:2012rk}, but certainly similar results can be obtained for other EOSs.  

\subsection{Conservative-to-primitive reconstruction}\label{c2p_reconsctruction}

Although the conservative variables are used as evolved variables, knowledge about the primitive variables is necessary to compute the fluxes, Eq.~\eqref{fluxterm}. Unfortunately, in contrast to transforming the primitive variables to the conservative variables through a simple analytical transformation, there is no closed analytical description to compute the primitive variables from the conservative ones. While there are numerous different possibilities for recovering the primitive variables, we will follow the discussion of Ref.~\cite{Thierfelder:2011yi} for ideal gas EOS and of Refs.~\cite{Galeazzi:2013mia, Kastaun:2020uxr} for the employed tabulated EOS. 

\subsubsection{Conservative-to-primitive recovery for an ideal gas}

For an ideal gas EOS, the pressure is given by
\begin{equation}\label{ideal_gas}
    p = (\Gamma -1)\rho \epsilon,
\end{equation}
where we use $\Gamma = 5/3$ in this article.
To compute the pressure $p$ from a set of conservative variables, we start with an initial guess $p^*$ and compute the following quantities: 
\begin{align}
    v^i &= \frac{S^i}{\rho h W^2} = \frac{S^i}{\tau + D + p^*},\label{velocity} \\
    W(p^*) &= \frac{1}{\sqrt{1 - v^2(p^*)}}, \\
    \rho(p^*) &= \frac{D}{W(p^*)}, \\
    \epsilon(p^*) &= \frac{\epsilon + \rho_0 W\left(1-W\right) + p^*\left(1-W^2\right)}{\rho W^2}, \\
    & = \frac{\epsilon + D\left(1-W(p^*)\right) + p^*\left(1-W^2(p^*)\right)}{DW(p^*)}.\label{internal_energy}
\end{align}
Once $\rho(p^*)$ and $\epsilon(p^*)$ are known, the ideal gas EOS in Eq.~\eqref{ideal_gas} allows us to compute the pressure $p(p^*)$. 
Following this procedure, we then iteratively update the initial guess $p^*$ until the residual \begin{equation}\label{residual}
    f(p^*) = p^* - p\left(\rho(p^*), \epsilon(p^*)\right),
\end{equation}
vanishes. For practical applications, Eq.~\eqref{residual} is solved with a root solver, e.g., a Newton-Raphson method. In this case, the update of the guess would be determined by
\begin{equation}
    p^{\rm new} = p^{\rm old} - \frac{f(p^{\rm old})}{f'(p^{\rm old})}
\end{equation}
where $p^{\rm new}$ and $p^{\rm old}$ are the pressure calculated at the current and the previous iteration through the loop, respectively. 
The derivative $f'(p)$ is given by
\begin{align}
    f'(p^*) &= 1 - \chi\frac{\partial \rho}{\partial p^*} - \kappa\frac{\partial\epsilon}{\partial p^*}, \\
    \frac{\partial \rho}{\partial p^*} &= \frac{DS^2}{(D + p^* + \tau)^2\sqrt{(D + p^* + \tau)^2} - S^2}, \\
    \frac{\partial\epsilon}{\partial p^*} &= \frac{p^* S^2}{D((D + p^* + \tau)^2 - S^2)^{3/2}},
\end{align}
where $\chi = \frac{\partial p}{\partial \rho} = (\Gamma-1)\epsilon$, $\kappa = \frac{\partial p}{\partial \epsilon} = (\Gamma-1)\rho$ in case of an ideal gas EOS, and $S=\sqrt{S^iS_i}$.

\subsubsection{Conservative-to-primitive recovery for tabulated EOSs}

In the case of a tabulated EOS, we follow the recovery method of~\cite{Galeazzi:2013mia, Kastaun:2020uxr}. Introducing the following quantities 
\begin{align}
    a &= \frac{p}{\rho(1 + \epsilon)}, & z &= Wv, \\
    q &= \frac{\tau}{D}, & r &= \frac{\sqrt{S^i S_i}}{D}, & k &= \frac{r}{1 + q},
\end{align}
one can see that
\begin{align}\label{redefined_vars_z}
    z &= \frac{r}{h}, \quad \rho = \frac{D}{W}, \quad  W = \sqrt{1 + z^2},\\
    \epsilon &= Wq - zr + W -1,\label{redefined_vars_eps} \\
    h &= (1 + \epsilon)(1 + a) = (W - zk )(1 + q)(1 + a),\label{redefined_vars_h}
\end{align}
holds. 
With the new variables introduced, the EOS is described by $a(\rho, \epsilon, p)$. 
In the following, we will describe only the necessary steps to recover the primitives here and refer readers to~\cite{Galeazzi:2013mia} for details, e.g., the constraints that the EOS should satisfy, the bounds for the conserved variables, and the existence of a (unique) solution within the bounds. 

For a given set of conserved variables, with the help of Eqs.~\eqref{redefined_vars_z}-\eqref{redefined_vars_h}, we can write the primitive variables as a function of $z$ with the definitions
\begin{align}
    \tilde{W}(z) &= \sqrt{1 + z^2}, \\
    \tilde{\rho}(z) &= \frac{D}{\tilde{W}(z)},  \label{W_tilde} \\
    \tilde{\epsilon}(z) &= \tilde{W}(z)q -zr + \frac{z^2}{1 + \tilde{W}(z)} \label{eps_tilde}.
\end{align}
Note that ($W -1$) has been replaced from Eq.~\eqref{redefined_vars_eps} with an equivalent expression $z^2/(1 + \tilde{W}(z))$, as the latter improves numerical accuracy when small velocities are encountered. 
The functions in Eqs.~\eqref{W_tilde} and \eqref{eps_tilde} can produce values for $(\tilde{\rho}, \tilde{\epsilon})$ that are outside the validity range of the EOS. 
To avoid this, the EOS function $a(\rho, \epsilon)$ is extended to $\mathbb{R}^2$ by 
\begin{align}
    \hat{\rho}(\rho) & \equiv {\rm max}({\rm min}(\rho^{\rm max}, \rho), \rho^{{\rm min}}), \\
    \hat{\epsilon}(\epsilon, \rho) & \equiv {\rm max}({\rm min}(\epsilon^{{\rm max}}(\hat{\rho}(\rho), \epsilon), \epsilon^{{\rm min}}(\hat{\rho}(\rho))), \\
    \hat{a}(\epsilon, \rho) & \equiv a(\hat{\rho}(\rho), \hat{\epsilon}(\epsilon,\rho)),
\end{align}
where $\epsilon^{{\rm min}}(\hat{\rho}(\rho))$ is computed at the minimum tabulated temperature, while $\epsilon^{\rm max}(\hat{\rho}(\rho), \epsilon)$ is computed initially at the maximum tabulated temperature and refined across successive iteration values of the temperature. $\rho^{\rm min}$ and $\rho^{\rm max}$ are the minimum and maximum values of $\rho$ from the EOS table, respectively. 
Then $\tilde{a}(z)$ and $\tilde{h}(z)$ are defined as follows
\begin{align}
    \tilde{a}(z) &= \hat{a}(\tilde{\rho}(z), \tilde{\epsilon}(z)), \\
    \tilde{h}(z) &= (1 + \tilde{\epsilon}(z))(1 + \tilde{a}(z)).\label{h_tilde}
\end{align}
From Eq.~\eqref{redefined_vars_z}, we then find the following master equation
\begin{equation}\label{master_eqn}
    f(z) = z - \frac{r}{\tilde{h}(z)}.
\end{equation}
The function $f(z)$ is well behaved for all values of $z$ and any root $z_0$ is confined to the interval $[z_-, z_+]$, with the bounds given by~\cite{Galeazzi:2013mia}
\begin{equation}\label{bounds for z Illinois}
    z_- = \frac{k/2}{\sqrt{1 - k^2/4}}, \qquad z_+ = \frac{k}{\sqrt{1 - k^2}}.
\end{equation}
Eq.~\eqref{master_eqn} can then be solved using bracketing root-finding methods. In our numerical-relativity code BAM~\cite{Thierfelder:2011yi,Gieg:2022mut,Schianchi:2023uky}, the Illinois method~\cite{Dowell1971} is used, which is a hybrid root-solver combining bisection, secant, and inverse quadratic interpolation methods.

\subsection{Neural networks}\label{section_NN}

In this work, we employ a supervised machine learning technique~\cite{russel2010, prince2023understanding} to predict the primitive variables from conservative ones. 
In particular, we use feedforward NNs~\cite{Schmidhuber_2015}, which form a computational circuit consisting of several layers of single units, referred to as neurons, that together define a function.
We restrict ourselves to fully connected NNs, where all neurons of adjacent layers are connected with tunable weights. 
Feedforward NNs often have several so-called hidden layers between the input and output layer.
As such, this class of machine learning methods is also known as deep learning~\cite{bengio2007scaling, lecun2015deep, goodfellow2016deep}. 
Once the NN receives an input, the value of a neuron in a subsequent layer is computed as a weighted sum of the previous layers' neurons, which is then passed to a non-linear activation function. The activation function outputs only when the weighted sum is greater than a certain threshold defined by the choice of activation function.
The network can be trained (i.e., adapting the connection weights) by providing example input-output pairs of the ground truth and minimizing a loss function that quantifies the difference between the NN prediction and the true output value. 
The minimization procedure relies on backpropagation~\cite{Rumelhart:1986gxv}, where gradients of the loss function are accumulated and used to adapt the weights with a certain step size, also referred to as the learning rate. 

Since NNs are universal function approximators~\cite{Hornik:1989yye}, our goal is to train NNs to approximate a map from the conservative variables to the primitive ones. 
The input of our NNs then consists of the conservatives $(D, |S|, \tau)$, while the output will be either the pressure $p$ or velocity-based quantity $z$. 
From this output, we can recover the other primitive variables following Eqs.~\eqref{velocity}-\eqref{internal_energy}. \\

For our implementation, we use the Fast Artificial Neural Network (FANN) library~\cite{Nissen:2003, fann}. 
We use the sigmoid activation $\sigma$ (with output range $[0,1]$) in the output layer and the symmetric sigmoid activation function $\sigma_{\rm sym}$ (with an output range of $[-1,1]$) in the hidden layers, which are defined as
\begin{align}
    \sigma(x) &= \frac{1}{1 + e^{-x}}, \\ \sigma_{\rm sym}(x) &= 2 \sigma(x) - 1 .
\end{align}
To train the network, we use the mean squared error (MSE) as loss function and adapt the weights using the resilience backpropagation algorithm~\cite{Ried_1994Rprop}.
Our training datasets contain 280,000 examples, while the test datasets, used to verify the generalization of the NNs to unseen data, have 20,000 examples. 
We have verified that increasing the training dataset does not significantly improve the performance of our networks.
The networks are trained for 2000 epochs.\footnote{Training is done on a machine with Intel(R) Core(TM) i7-10610U CPU @ 1.80GHz, which has 8 logical CPUs and x86\_64 architecture. The time taken to train the network is $\sim 21$ minutes and $\sim 5$ minutes for the ideal gas EOS and SFHo EOS, respectively.} 

The precise details of our architectures and the mechanism to construct training data vary for the two EOSs considered in this work.
In the following, we describe the NN architectures and training procedures for each EOS in more detail.

\subsubsection{Neural network for the ideal gas EOS conservative-to-primitive recovery}

To construct the training data for the ideal gas EOS, we draw values for $\rho$ and $\epsilon$ from a log-uniform distribution with  $\rho \in (10^{-14},10^{-2})$, $\epsilon \in (10^{-12}, 10^2)$. The velocity $|v|$ is sampled uniformly between $0.001$ and $0.8$.
From these samples, the pressure is computed using Eq.~\eqref{ideal_gas}, after which the corresponding conservative variables $(D, |S|, \tau)$ are computed with Eqs.~\eqref{cons_D}-\eqref{cons_tau}. 
This gives us the conserved variables within the range $(1.0\times10^{-14}, 0.02)$ for $D$, $(1.3\times10^{-17}, 2.6)$ for $|S|$, and $(2.83\times10^{-20}, 2.79)$ for $\tau$. Because of such large input data ranges, we have further preprocessed the data. We convert the data into $log$ scale to minimize the ranges. To further reduce the range, we scale the data into a $[0, 1]$ range using the ${\rm MinMaxScaler}$ from scikit-learn~\cite{scikit-learn}. When implemented in BAM, we have to do the above scaling of conservative variables before they are given as input to the NN, and when the prediction is received from the NN, we have to invert back to the normal scale that we have started with. 

The NN architecture has one input layer with 3 neurons for the values $(D, |S|, \tau)$, followed by 3 hidden layers with 20, 20, and 15 neurons, respectively. 
The output layer has one neuron, with the pressure $p$ being the output variable.
The learning rate is set to $0.7$, and the validation error of $3 \times 10^{-5}$ is achieved at the end of the training. 

In the following, we will see that the constructed NN will be sufficiently accurate for special-relativistic shocktube tests (Sec.~\ref{sec:schocktube}) but will not be sufficient for general-relativistic simulations. In such cases, as we will discuss later, the network will provide initial guesses for the root solver; cf.~Sec.~\ref{TOV_ideal}.

\subsubsection{Hybrid conservative-to-primitive recovery for tabulated EOS}
\label{tabulated_EOS}

For the tabulated SFHo EOS, the training dataset has been constructed as follows.
First, we construct a 4-dimensional parameter space $(\rho, T, Y_e, |v|)$, where $\rho$, $T$, and $Y_e$ are taken from the EOS table by sampling them log-uniformly within the provided EOS ranges. Then computing the conserved variables with the distributed primitives, we end up with the ranges $(2.68\times10^{-15}, 0.14)$ for $D$, $(4.81\times10^{-19}, 19.12)$ for $|S|$, and $(3.19\times10^{-17}, 18.98)$ for $\tau$. 
The velocity $|v|$ is sampled independently from a uniform distribution between $0$ and $0.99$. 
For each set of primitive variables, we compute $z = W|v|$ and the conservative variables $(D,|S|,\tau)$ (Eqs.~\eqref{cons_D}-\eqref{cons_tau}). Regarding the preprocessing of the data, we follow the same steps as for the ideal gas EOS.  

The NN architecture consists of an input layer with $3$ neurons for the conservative variables $(D, |S|, \tau)$ and $4$ hidden layers, each consisting of $3$ neurons.
The output layer consists of a single neuron representing the variable $z$. 
The input does not consider $Y_e$, since the variable $z$ does not significantly depend on $Y_e$ for the SFHo EOS considered here. 
While this design choice potentially reduces the accuracy of the network, we found that it results in smaller networks that are cheaper to evaluate and still achieve high enough accuracy for our simulations if combined within a hybrid approach with more traditional methods (see below). The learning rate is set to $0.001$, and we obtain a validation error of $6.0\times10^{-3}$ after the training. \\

Compared to the ideal gas EOS, approximating the conservative-to-primitive routine for a realistic, tabulated EOS is challenging. 
As a result, we have found that the NN cannot be used directly as a replacement for the conservative-to-primitive routine.
Reducing this error of the NN predictions requires a more complex NN architecture, which would then dominate the execution time. 
Therefore, we propose a hybrid conservative-to-primitive recovery scheme for tabulated EOS.
In particular, we use the NN output as an initial guess for the Illinois root-finding method~\cite{dahlquist1974numerical, FordImprovedAO} implemented in BAM. 
Since the root-finding algorithm ensures that we achieve the numerical accuracy required by simulations, we opt for a simple NN architecture that is efficient to evaluate. The initial guess coming from the NN is then instead used to constrain the bounds employed in the Illinois scheme, cf.~Eq.~\eqref{bounds for z Illinois}.
After computing the NN prediction $z_{\rm NN}$ for the root, our algorithm verifies if $z_{\rm NN} \in [z_-, z_+]$. 
If this is not the case, the algorithm defaults to the Illinois method.
If, however, $z_{\rm NN}$ lies within the bounds, it further checks the sign of the function value $f(z_{\rm NN})$. 
Depending on the sign, the bounds are adjusted. 
After that, the Illinois algorithm is executed with the adjusted bounds.

\subsection{Hydrodynamical simulations with BAM}

For our hydrodynamical simulations, we are using the BAM code~\cite{Bruegmann:1997uc,Bruegmann:2006ulg,Thierfelder:2011yi, Dietrich:2015iva,Bernuzzi:2016pie,Neuweiler:2024jae, Schianchi:2023uky, Gieg:2022mut}. 
BAM uses an adaptive mesh refinement method based on multiple levels, labeled from $l = 0$ (coarsest) to $l = L-1$ (finest). Each level can consist of a number of refinement boxes, where for levels $l > l_{\rm mv}$, the refinement boxes can move dynamically during the simulation to follow the motion of the compact objects. 
The grid spacing at each level follows a 2:1 refinement strategy, meaning the spacing at level $l$ is half of the spacing at level $l-1$. All grid points are cell-centered and staggered to avoid singularities at the origin. 

The time evolution is handled using a Berger-Oliger~\cite{Berger:1984zza} scheme for local-time stepping and the Berger-Colella~\cite{Berger1989JCoPh} scheme to ensure flux conservation across refinement boundaries. For solving spacetime evolution, we use here the Z4c formulation~\cite{Bernuzzi:2009ex, Hilditch:2012fp}, discretized with a fourth-order finite difference scheme. The lapse function is evolved with $1 +{\rm log}$ slicing~\cite{Bona:1994dr}, and the shift vector follows the gamma-driver conditions~\cite{Alcubierre:2002kk}.

For the hydrodynamic equations, cf.~Sec.~\ref{sec:GRHD_equations}, we reconstruct the primitive variables employed for the flux computation using the WENOZ~\cite{Borges2008JCoPh, Bernuzzi:2016pie} method together with the HLL Riemann solver~\cite{Amiram:2006zjz}. 

\section{Method Validation}
\label{sec:validation}

\subsection{Special-relativistic shocktubes for the ideal gas EOS}
\label{sec:schocktube}

\begin{figure}[t]
    \centering
    \includegraphics[width=\linewidth]{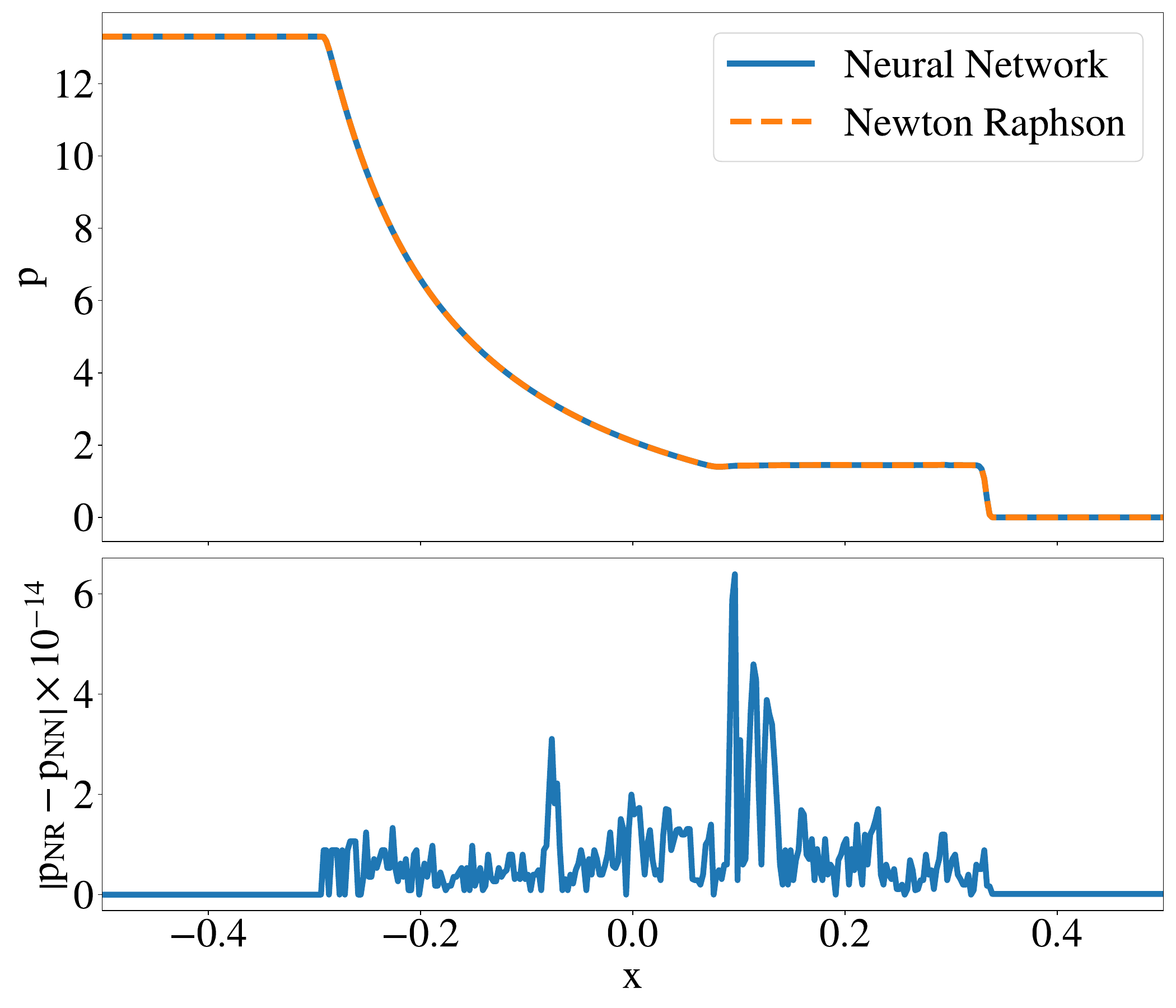}
    \caption{Top panel shows the evolution of pressure when the interface is removed. Comparison between the NN and the root solver at $t=0.4$. The bottom panel shows the absolute difference between the pressure computed by the Newton-Raphson method ($p_{\rm NR}$) and NN ($p_{\rm NN}$).}
    \label{fig:shocktube_p}
\end{figure}

Inspired by the works of Refs.~\cite{Dieselhorst:2021zet,Galeazzi:2013mia, Kacmaz:2024fwa}, we start by testing the applicability of our implementation with a standard shocktube test~\cite{Marti1996, Donat1998, Font:1998hf}. 
The simulation is set up according to~\cite{Font:1998hf}, assuming an ideal gas EOS with $\Gamma=5/3$. 
The integration domain extends from $x_L=-0.5$ to $x_R=0.5$, and the initial state of the system is specified by
\begin{align*}
    p_L &= 13.3, & p_R &= 0.66\cdot10^{-6}, \\ 
    \rho_L &= 10, & \rho_R &= 1.
\end{align*}
For this test, the traditional conservative-to-primitive transformation is fully replaced by the predictions from the NN. 
Figure~\ref{fig:shocktube_p} demonstrates that the NN can replace root solvers in terms of accuracy and speed.
The maximum absolute error between the root solver and the NN for our prediction of $p$ is $6.395 \times 10^{-14}$, and the average error is $1.111 \times 10^{-15}$, which is sufficient to use the NN prediction directly for dynamical simulations.

The time\footnote{The simulation was performed on an in-house workstation with Intel(R) Xeon(R) Gold 6248R CPU @ 3.00GHz, which has 96 logical CPUs and x86\_64 architecture. Only one CPU is used to simulate the shocktube test.} taken for the simulation (averaged over 10 simulations) is $13.490\ \rm s$, whereas, when using a root solver, the simulation was completed in $12.714\ \rm s$. Given the difference of $\sim0.7\ \rm s$ for both methods, it seems natural to potentially reduce the size of the network, reducing the accuracy until a still acceptable level, for instance, by pruning our NNs~\cite{OptimalBrainDamage, OptimalBrainSurgeon}. 

\subsection{Single TOV test with ideal gas EOS}\label{TOV_ideal}

\begin{figure}
    \centering
    \includegraphics[width=\linewidth]{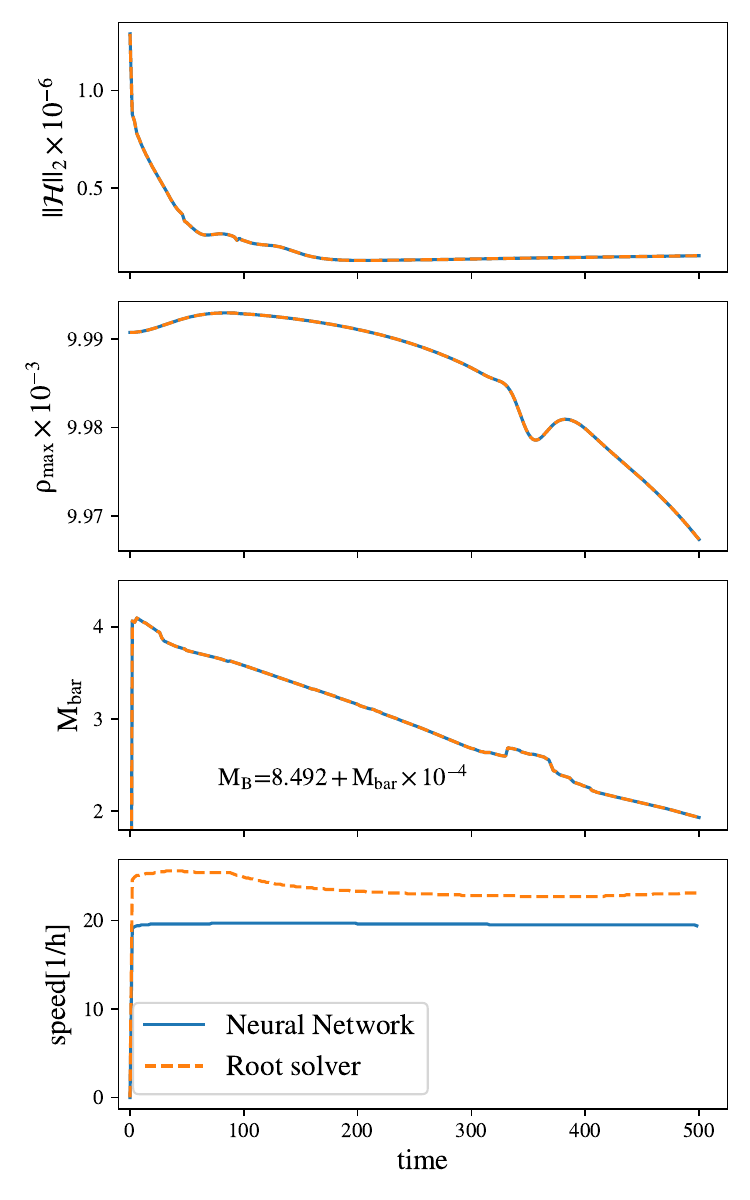}
    \caption{Simulation of single TOV star with ideal gas EOS. The top panel shows the Hamiltonian constraint $||\mathcal{H}||_2$, the second panel shows the central density, the third panel shows the total baryonic mass, and the bottom panel shows the comparison of speed between the traditional root solver and the NN-based approach. However, we emphasize that in this test -- in contrast to Fig.~\ref{fig:shocktube_p} -- the NN prediction simply provides an initial guess for a Newton-Raphson root solver, since purely using the NN did not lead to long-term stable simulations.}
    \label{fig:tov_ideal_gas}
\end{figure}

As the next test case, we will consider a spherically symmetric star described with an ideal gas EOS with $\Gamma=5/3$. We obtain the initial configurations by solving the Tolmann-Oppenheimer Volkoff Equations~\cite{Tolman:1939jz, Oppenheimer:1939ne}. Because of the chosen EOS, the star has a very large radius of $44.5$, and should not be considered as a realistic test, but rather to check the consistency of the implemented method. 
The simulation employs five refinement levels with a grid spacing of $0.5$ in the finest level. To save computational costs, we employ octant symmetry, i.e., just one octant of the full grid for which we would have 256 grid points per direction.  

As for the shocktube test, we employ the Newton-Raphson as our default root-solver option and compare it with the performance of the NN conservative-to-primitive routine. However, for this particular case, we find that the NN prediction is not always accurate enough, and therefore, we decided to use the NN prediction as our initial guess for the Newton-Raphson root solver. 

Our simulation results are shown in Fig.~\ref{fig:tov_ideal_gas} and show agreement between both methods regarding the Hamiltonian constraint, the central density, and the baryonic mass (top three panels). Considering the simulation speed (bottom panel), we find that our NN method is slower than the traditional Newton-Raphson root solver. Clearly, this is not surprising, as our NN prediction here only serves as an initial guess for the root solver. However, this leads to two immediate conclusions: (i) although NN predictions can be accurate enough in specialized special-relativistic shocktubes, as shown in previous works in the literature~\cite{Dieselhorst:2021zet}, they might fail for more complicated scenarios; (ii) since potentially one always has to use a hybrid approach in which a root solver simply takes the NN prediction as an initial guess, it might be better to use a simple NN architecture to reduce computational costs. Both aspects have motivated our design choices for the following studies using a tabulated EOS. 

\subsection{Single TOV tests with a tabulated EOS}
\label{results: single TOV with tabulated EOS}

As our next test, we consider a single TOV star but with the tabulated EOS SFHo. We employ a grid setup consisting of three refinement levels with a grid spacing of 0.233 in the finest level. 
We employ 96 points per direction and use octant symmetry to reduce computational costs. 
For this test, the NN can give accurate predictions for most cases, but it fails for some values, e.g., near the extremes or outside of the trained data. So, our implementation is as follows. We replace the root solver with NN, but when NN fails, we default to the root solver. This hybrid method increases the speed of this simulation while ensuring its stability.

\begin{figure}[htpb]
    \centering
    \includegraphics[width=1\linewidth]{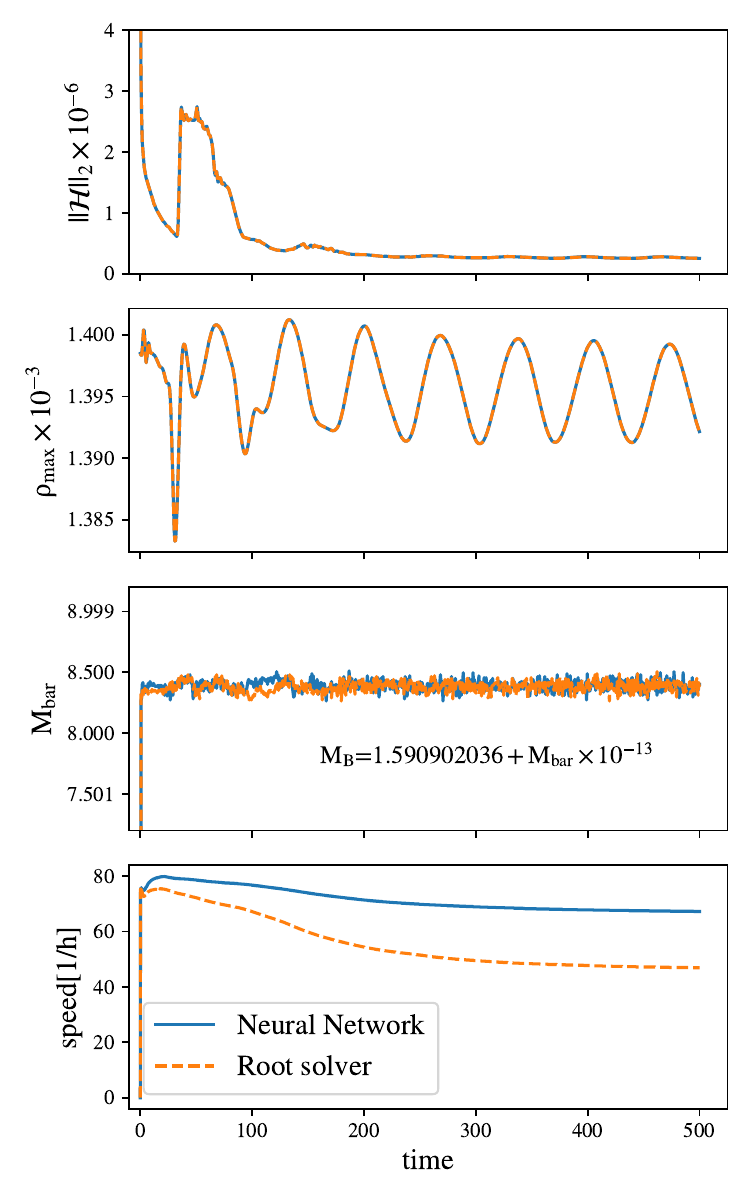}
    \caption{Simulation of a single TOV star employing the SFHo EOS as discussed in the main text. The panels show the Hamiltonian constraint $||\mathcal{H}||_2$, the central density, and the total baryonic mass, as well as the speed of the simulations (from top to bottom). }
    \label{fig:tov_plots}
\end{figure}

Fig.~\ref{fig:tov_plots} compares the results of the simulation employing our algorithm with that using the traditional root solver. 
The stability of the new method is confirmed by the agreement between the Hamiltonian norm $||\mathcal{H}||_2$ and the central density $\rho_{\rm max}$, in the first and second panel of Fig.~\ref{fig:tov_plots}.
In the third panel, we show the total baryonic mass $M_{B}$, which differs from the Illinois method only at the order of $10^{-14}$. 
The bottom panel shows a direct comparison of the speed of simulation for the different methods. 
One can see that by employing the NN, the speed of the simulation increases by up to 30\%\footnote{This test was performed on a machine with Intel(R) Core(TM) i9-10900K CPU @ 3.70GHz, that has 20 logical CPUs and x86\_64 architecture. 10 logical CPUs were used for the simulation with OpenMP parallelization.}.
Therefore, employing our proposed method holds promise to accelerate more complex simulations, as we explore in the next section.

\section{Simulating Binary Neutron Stars}
\label{sec:BNS}

\begin{figure*}[t]
    \centering
    \includegraphics[width=\linewidth]{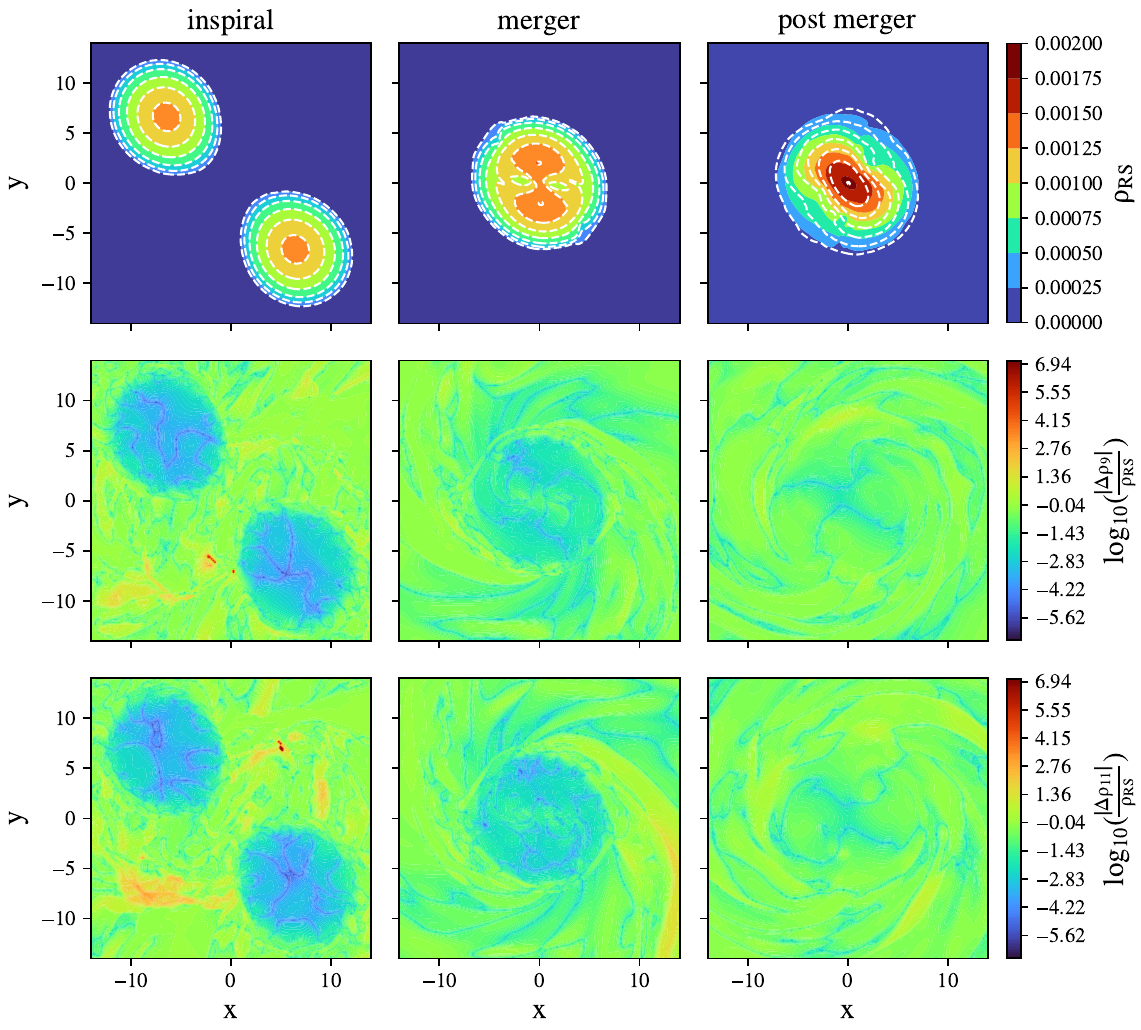}
    \caption{Comparison of the density $\rho$ between the root solver and our deep-learning algorithm for conservative-to-primitive conversion in the simulation of a BNS merger. The top row shows the density $\rho$ computed by employing the root solver ($\rho_{\rm{RS}}$) at three different times during the simulation. We overlay the same density regions with white contours for the NN-informed simulation (using a conservative-to-primitive accuracy threshold of $10^{-11})$. The middle, respectively bottom row shows the relative difference in density $\frac{\Delta\rho}{\rho}$, compared to our deep-learning method, when using an error threshold of $10^{-9}$, respectively $10^{-11}$ for the root solver. }
    \label{fig:bns_density}
\end{figure*}

As our final test to check the validity of the proposed method, we want to investigate how the modified conservative-to-primitive routine performs during BNS simulations.  For this purpose, we simulate an equal mass binary in which the stars have a gravitational mass of $M_1=M_2=1.28$, a baryonic mass of $M_{b,1}=M_{b,2}=1.4$. As in the previous test, we employ the SFHo EOS. 

The simulations use a grid consisting of seven refinement levels with 192 points per direction on the coarser levels. The two finest levels are split into two boxes with 96 points per direction. The boxes can follow the movement of the stars to ensure high resolution around the regions with the strongest spacetime curvature. The code uses a 2:1 refinement strategy, in which each coarser level has a doubled grid spacing. The resolution in the finest level is about 0.168.

Similar to our single TOV test with the tabulated EOS of Sec.~\ref{results: single TOV with tabulated EOS}, we employ the hybrid algorithm described in Sec.~\ref{tabulated_EOS} and use the NN prediction as an initial guess for the Illinois algorithm. 

\begin{figure}[t]
    \centering
    \includegraphics[width=\linewidth]{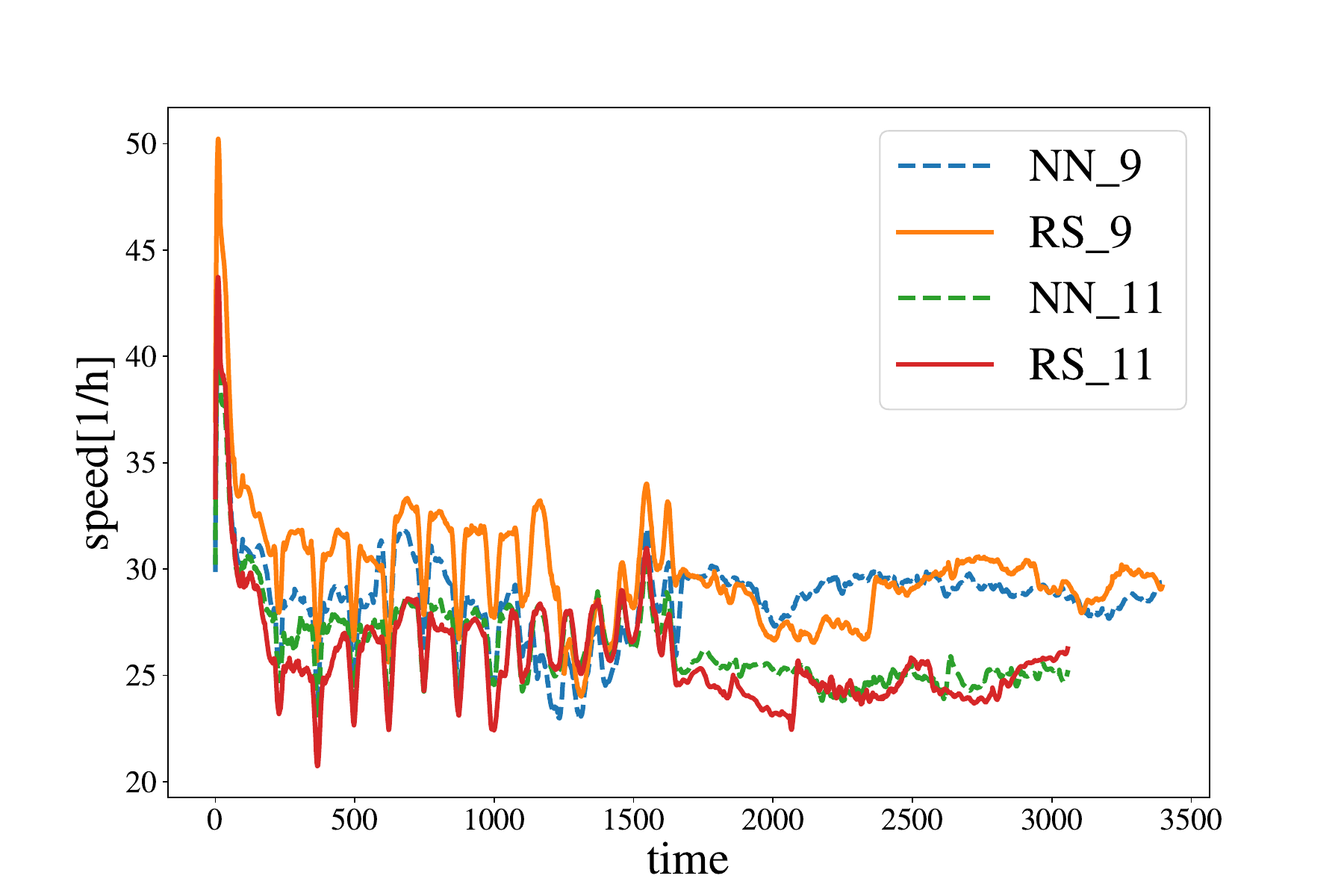}
    \caption{Comparison of speed for the simulation of a BNS merger with two different error threshold values. NN represents that the NN has been employed in conservative-to-primitive computation, while RS  represents that the root solver alone has been used for the conservative-to-primitive computation. The numbers indicate the error threshold, which is set to either $10^{-9}$ or $10^{-11}$.}
    \label{fig:speed_bns}
\end{figure}

In Fig.~\ref{fig:bns_density}, we show the result of our dynamical simulation, where the top row shows the density profile at three different times computed with the standard root solver ($\rho_{\rm{RS}}$) used for the conservative-to-primitive recovery. In addition, we overlay the same density contours when using the NN-informed conservative-to-primitive routines, and we find a very good agreement. 
The middle and bottom panels show the absolute difference in density when instead employing our deep-learning based algorithm. 
Our comparison considers two different accuracy thresholds for the conservative-to-primitive recovery, namely $10^{-9}$ (middle panel) and $10^{-11}$ (bottom panel), to assess the influence of the threshold on the overall runtime. While the results agree for both thresholds, using a smaller threshold for the conservative-to-primitive routine reduces the differences between the two methods. This simulation can be considered as a proof of principle that our proposed deep-learning method can be used during dynamical simulations of neutron star spacetimes. 

In Fig.~\ref{fig:speed_bns}, we compare the speed of our BNS simulations. 
These simulations are performed on the HAWK System of the H\"ochstleistungsrechenzentrum Stuttgart using AMD EPYC 7742 CPUS.
We have run the simulations on one compute node employing 8 MPI tasks with each 16 OpenMP threads per MPI-task. 

For the error threshold of $10^{-9}$, we find that the Illinois method has an average speed of $52/h$. 
In contrast, the average reduces to $49.3/h$ when NN predictions were used as input for the initial guess. 
For the accuracy error threshold of $10^{-11}$, we obtain an average speed of $48.0/h$ for the Illinois method and $50.4/h$ with the NN input. 
However, it is clearly visible from Fig.~\ref{fig:speed_bns} that in real applications, the variation in the speed, e.g., due to I/O, communication, and the overall load on the system, strongly influences the speed. 
Hence, the speedup or slowdown that the new conservative-to-primitive method achieves is subdominant, and one can conclude that similar performance is obtained within realistic scenarios. 
Our tests\footnote{These tests considered the early-inspiral stage for simplicity, but confirmed our results overall.} also confirm this by employing different HPC systems and grid resolutions. 

\section{Conclusion}

General-relativistic hydrodynamics simulations of compact binary mergers are generally connected to high computational costs. For this reason, several attempts have been made to reduce the computational footprint of such simulations. In this regard, Dieselhorst et al.~\cite{Dieselhorst:2021zet} and, later, Kacmaz et al.~\cite{Kacmaz:2024fwa} investigated the possibility of using NNs to speed up the required conservative-to-primitive recovery during hydrodynamical simulations. 
However, none of these works employed NNs in an actual simulation of a BNS merger, which was the main aim of the present work. 

While we have found that the usage of NNs achieves a sufficiently high accuracy to be directly employed in simple tests, during the simulation of neutron star spacetimes, the recovery is not accurate enough to avoid the use of a root-finding algorithm. We expect that more complex networks could overcome this issue, at the cost of a more expensive NN to evaluate. Therefore, we have employed a hybrid approach in which we use NNs to improve the initial guesses handed to a root-solver routine. Since the exact speed depends on the resolution, the requested accuracy threshold for the conservative-to-primitive routine, as well as the computational setting, i.e., the number of nodes and the employed MPI/OpenMP settings, our results should only be considered as estimates.  

It is worth highlighting some of the existing limitations of our work. First, our study was restricted to the usage of two different EOSs, one rather simple ideal gas EOS with $\Gamma=5/3$ and one tabulated EOSs based on the SFHo~\cite{Steiner:2012rk} EOS. Although there is no reason to expect that other EOSs could not be used, it might be that the presence of strong phase transitions or other particular features would make the training harder. In general, for each different EOS, we would require a newly trained NN. While the training of the NN adds only a small amount of extra computational costs (see section \ref{section_NN}), it certainly adds another layer of complexity, in particular, for the studies that involve a large number of EOSs. 

Second, we expect that further fine-tuning of the trained NN might further reduce the computational footprint. On this behalf, we think that, with the upcoming GPU-based generation of numerical-relativity codes, NNs will become more competitive with respect to traditional root finders, due to their simpler flow chart.

Third, another possible speedup could be obtained by connecting the NN prediction with a Newton-Raphson method because of its higher convergence order compared to the Illinois method. However, this approach would require us to know accurate derivatives of the pressure. In principle, as shown by Dieselhorst et al.~\cite {Dieselhorst:2021zet}, one could also train an NN to predict these derivatives. However, this also increases the complexity of the overall method and would require detailed tests. 

\section*{Acknowledgement}

The authors thank Michele Mattei, Pouyan Salehi, and Ka Wa Tsang for discussions in a very early stage of the project. 
This work was partially funded by the European Union (ERC, SMArt, 101076369). Views and opinions expressed are those of the authors only and do not necessarily reflect those of the European Union or the European Research Council. Neither the European Union nor the granting authority can be held responsible for them. 
T.W. acknowledges funding through NWO under grant number OCENW.XL21.XL21.038.
F.S. acknowledges funding through the grant PID2022-138963NB-I00 funded by MCIN/AEI/10.13039/501100011033/FEDER, UE

The simulations were performed on the national supercomputer HPE Apollo Hawk at the High Performance Computing (HPC) Center Stuttgart (HLRS) under the grant number GWanalysis/44189, and at the DFG-funded research cluster (INST 336/173-1; project number: 502227537) jarvis.

\textbf{Conflict of Interest Statement} The authors have no competing interests to declare that are relevant to the content of this article.

\textbf{Data Availability Statement} Data will be made available upon reasonable request from the corresponding author.

\textbf{Code Availability Statement} Code used to produce the results of this article is proprietary and cannot be made available.

\printbibliography

\end{document}